\documentclass[twocolumn,english,showpacs,preprintnumbers,amsmath,amssymb,floatfix,nofootinbib]{revtex4-1}
\usepackage[T1]{fontenc}
\usepackage[latin9]{inputenc}
\usepackage{color}
\usepackage{array}
\usepackage{amstext}
\usepackage{graphicx}
\usepackage{esint}
\usepackage{rotating}
\usepackage[font=small,labelfont=bf]{caption}
\usepackage{xcolor}
\usepackage[colorlinks = true,
linkcolor = magenta,
urlcolor  = blue,
citecolor = red,
anchorcolor = blue]{hyperref}

\def\as{\relax\ifmmode \alpha_s\else{$ \alpha_s${ }}\fi}  %%%%%%%%%
\def\abar{\relax\ifmmode{\bar{a}}\else{$\bar{a}${ }}\fi}  %%%%%%%%%
\def\be{\begin{equation}}
\def\ee{\end{equation}}
\def\ba{\begin{eqnarray}}
\def\ea{\end{eqnarray}}
\makeatletter

\@ifundefined{textcolor}{}
{%
 \definecolor{BLACK}{gray}{0}
 \definecolor{WHITE}{gray}{1}
 \definecolor{RED}{rgb}{1,0,0}
 \definecolor{GREEN}{rgb}{0,1,0}
 \definecolor{BLUE}{rgb}{0,0,1}
 \definecolor{CYAN}{cmyk}{1,0,0,0}
 \definecolor{MAGENTA}{cmyk}{0,1,0,0}
 \definecolor{YELLOW}{cmyk}{0,0,1,0}
 }

\@ifundefined{definecolor}
 {\usepackage{color}}{}
\@ifundefined{definecolor}
 {\usepackage{color}}{}
\makeatother
\usepackage{babel}

\begin{document}

%%%%%%%%%%%%%%%%%%%%%%%%%%%%%%

\title {\textcolor[rgb]{0.00,0.00,0.00}{Comparing optimized renormalization schemes for QCD observables}}
%Analysing the McKeon  and et al. approach in comparison with the approach of complete renormalization grprovement}}

\author {M.~Akrami}
\email{Akrami@stu.yazd.ac.ir}

\author{ A.~Mirjalili}
\email{Corresponding author:A.Mirjalili@yazd.ac.ir}

\affiliation {Physics Department, Yazd University, P.O.Box 89195-741, Yazd, Iran}

\date{\today}

%
%%%%%%%%%%%%%%%%%%%%%%%%%%%%%%%%%%%%%%%%%%%%%%%%%%%%%%%%%%%%%%%%%%%%%%%%%%%%%%%%%%%%%%%%%%%%%%%%%%%%%%%
\begin{abstract}\label{abstract}
In this article based on  the McKeon  ${\it et\; al.}$ approach, it can be shown that the renormalization group equation  while it is related to the
radiatively mass scale $\mu$, operates to do a summation over  QCD perturbative  terms.  Employing the full QCD $\beta$-function within this summation,
all logarithmic corrections can be  presented as the log-independent contributions.
In  another step of  this approach , the renormalization scheme dependence for QCD observable which is characterised by Stevenson,  can be examined by
specifying the  renormalization scheme in which the  $\beta$-function beyond two-loop orders is utilized. In this regard two choices of renormalization
schemes would be exposed. In one of them the QCD observable is expressed while involves two powers of running coupling constant such that the
perturbative series contains just two terms.  In another choice the perturbative series expansion is written as an infinite series in terms of
two-loop
running coupling which can be presented by Lambert $W$-function. In both cases the QCD observable involves parameters which are renormalization scheme
invariant  and the coupling constant which is independent of renormalization scale.
We then consider the other approach which is called the complete renormalization group improvement (CORGI). In this approach, using the self consistency
principle  it is possible to reconstruct the conventional perturbative series in terms of  scheme invariant quantities and the coupling constant as a
function of Lambert-W function. It should be noted while in the renormalization group summation method of McKeon ${\it et\; al.}$,
scheme dependence of observables is investigated separately from their scale dependence,
in CORGI approach,
through the principle of self consistency,
both scale and scheme parameters are utilized together. In continuation we examine numerically these two approaches, considering two QCD observable. The first one is $R_{e^+e^-}$ ratio, investigated at three different colliding  energies
 and  the second observable is Higgs decay width to gluon-gluon. Comparison for $R_{e^+e^-}$ with the available  experimental is done. The results based on the McKeon ${\it et\; al.}$ approach are in better agreement with the experimental data.
\end{abstract}
\maketitle
\section{Introduction}\label{Introduction}
There are different approaches to optimize  the QCD observable. One of them is called principle of maximum conformality (PMC) which was intimated by
S.J.Brodsky and collaborators \cite{pmc11,pmc12,pmc14,pmc15,pmc16}. In this approach the  perturbative series is divided into two conformal and
non-conformal parts. The conformal
part does  involve the terms which are independent on the employed scheme. The non-conformal parts are absorbed into the coupling constant by choosing
proper
scales. During the absorbtion, the renormalization  scales would be fixed and determined and final result is such that the conformal series to be
independent of the used scheme and scales. The division of the series can be done, using an auxiliary scheme that is called $R_\delta$ scheme
\cite{pmc12}.
In connection to the PMC approach  some other research activities have been done which can be found in \cite{other1, other2}.

Another approach to optimize QCD observable is called complete renormalization group improvement (CORGI) \cite{cj-1}. In this approach, using self
consistency
principle, it is possible to write each expansion coefficient at high order, in terms of coefficients of lower orders and an invariant scheme independent
terms which are unknown at that high order. If one resumes the contribution of each expansion terms at lower orders  then it can be shown that the
result would be scale and scheme independent. This provides an opportunity to reconstruct the initial  perturbative series in terms of scale and scheme
independent
terms and also coupling constant which are determined at the physical energy scales and consequently the scale ambiguity is removed.

There might be some connections between the PMC and CORGI approaches which have been discussed in \cite{PRD2019-AM}. The most important one is that the
predictable terms in CORGI approach can be assigned  to the  non-conformal  part since the predictable terms are scale and scheme dependence while
unpredictable terms can be related to the conformal part of PMC which is scheme independent. For more details, see \cite{PRD2019-AM}.

As a third approach which has been initiated by D.~C.~McKeon and his collaborators and we call it McKeon  ${\it et\; al.}$ approach,  it can be shown that
upon resummation of all perturbative series terms to all
orders, using the full QCD-$\beta$-function, the result would be scale independent and just depends on the physical energy scale, $Q$, without any
ambiguity.
In continuation using the scheme dependence of coupling constant, it is shown how the perturbative coefficients up to specified high order, can be written
in terms of coefficients at lower orders toward  the lowest order coefficient  \cite{mac1,mac2,mac3,mac4,mac5}. This
stage is like the one in CORGI approach but with one little difference. In CORGI approach the perturbative coefficients at specified high order are
obtained in
terms of coefficients at  lower orders based on self consistency principle. In this regard, it is required to obtain the derivative of
perturbative coefficients not only with respect to the scheme parameters, but also with respect to renormalization scale which is casted in terms of
$\tau$
variable, see \cite{Maxwell-Mir2000}. Therefore, in comparison with McKeon  ${\it et\; al.}$ approach, the result of CORGI approach for perturbative
coefficient at a specified
high order, not only contains the coefficients at lower orders, but also involves more extra terms. This occurs, as we refereed above,
because the separation between scale and scheme dependence does not happen during the optimizing
procedure in CORGI approach while it has been supplied by the utilized approach of Mackeon  ${\it et\; al.}$\\
We are know planning to discuss in more details the differences and  similarities between CORGI and  Mackeon ${\it et\; al.}$ approach and to investigate
which
one has more preference. We specially examine this procedure, considering two QCD observable: one is the ratio of electron positron annihilation to
hadrons
and the other is the Higgs decay width to gluon-gluon. But at first we need to review the basic concepts of theses two approaches which it is done  in the
further sections.

The organization of this paper consists of the following section. In Sec.{\ref{rev}} basic concepts of McKeon   $\it et\; al.$  approach is reviewed. This
section contains two parts. At first part  the renormalization group summation  is  described which is done in Subsec.$\ref{AA}$. The scheme
characterizing
of QCD observable is done in Subsect.$\ref{B}$.
Further on we are going to examine the  Mackeon  ${\it et\; al.}$ approach  and the CORGI one as well  for some QCD observable which are done in
Sec.$\ref{exam1}$. At first we inspect $R_{{e^+}(e^-)}$ ratio in Subsect.$\ref{exam1A}$ and Subsect.$\ref{exam1B}$, considering McKeon   $\it et\; al.$
approach and CORGI one respectively.
We then employ the McKeon   $\it et\; al.$  approach but for Higgs decay width to gluon-gluon in Sec.$\ref{exam2}$ where we first there indicates  how to
do the RG summation. The scheme dependence for this observable is done in Subsect.$\ref{exam2A}$. Numerical investigation for this observable in McKeon
$\it et\; al.$  approach is presented in Subsect.$\ref{exam2B}$. This observable is considered numerically as well in CORGI approach in Sec.{\ref{cor}}.
Finally we give our conclusion in Sec.{\ref{con}}
\section{An overview on McKeon ${\it et\; al.}$ approach \label{rev}}
Here we give a brief description to the method, initiated  by McKeon   $\it et\; al.$ on renormalization group summation. Most of the relations in this
section and the further ones are rephrased from \cite{mac1,mac2}. We are inevitable to resort them in order the readers can follow easily and precisely
the
concerned subject. As an example how the renormalization group summation has been done, the QCD observable for $e^{+}e^{-}$ annihilation is examined. In
following, it is shown how  by achieving a recurrence relation between perturbative terms, the result of summation on all perturbative terms would be
independent of renormalization scale  as an unphysical parameter.

The cross section of $e^{+}e^{-}$ annihilation into hadrons, after normalizing it to $\mu^{+}\mu^{-}$ pair production, can be written as:
\be
{{R}_{{{e}^{+}}{{e}^{-}}}}=(3\sum\limits_{i}{q_{i}^{2})[1+R]}\label{per}\;.
\ee
The expansion of $R$ in terms of coupling constant $a$ has the  following form:
\be
R=\sum\limits_{n=0}^{\infty }{{{r}_{n}}{{a}^{n+1}}\,\,\,\,,\,\,\,\,\,\,\,{{r}_{0}}=1}\label{R}\;.
\ee
The coefficient of $r_{n}$, arising from contribution of Feynman diagrams to the observable $R$ can be represented by~\cite{mac1}:
\be
{{r}_{n}}=\sum\limits_{m=0}^{n}{{{T}_{nm}}{{L}^{m}}}\label{rn}\;.
\ee
In this equation $T_{00}=1$ and $L=b\ln (\frac{\mu }{Q})$ where $\mu$ is renormalisation scale while $Q$ is the physical center of mass energy in $e^{+}$
and $e^-$ collision .

Since $R$ is independent of renormalization scale
$\mu$, so the renormalization group  equation (RGE) implies that:
\be
\mu \frac{d}{d\mu }R=(\frac{\partial }{\partial \mu }+\beta (a)\frac{\partial }{\partial a})R=0\label{RG} \;.
\ee
The $\mu$-dependence of the coupling is governed by the QCD
$\beta$-function equation \cite{beta}:
\be
\beta (a)=\mu \frac{\partial a}{\partial \mu
}=-b{{a}^{2}}(1+ca+{{c}_{2}}{{a}^{2}}+{{c}_{3}}{{a}^{3}}+...)\label{beta}\;.
\ee
Here $b = (33-2N_{f} )/6$ and $c = (153-19N_{f} )/12b$ are renormalization scheme (RS) invariant where $N_f$ is denoting to the number of active quark
flavor.
The higher coefficients $c_{2},c_{3},...$ serve to label the RS dependence \cite{steven}.

To indicate the coefficients $b$ and $c$ in Eq.(\ref{beta}) are scheme independent, one can consider two
different coupling $a$ and $a^{*}$ at two different scheme, such that \cite{mac2}:
\be
{{a}^{*}}=a+{{x}_{2}}{{a}^{2}}+{{x}_{3}}{{a}^{3}}+...\;,\label{starr}
\ee
therefor
\be
{{\beta }^{*}}({{a}^{*}})=\mu \frac{\partial {{a}^{*}}}{\partial \mu }=\beta (a)\frac{\partial
{{a}^{*}}}{\partial a}\;.
\ee
Then one can show:
\begin{align}
& {{\beta }^{*}}({{a}^{*}})=-{{b}^{*}}{{a}^{*}}^{2}[1+{{c}^{*}}{{a}^{*}}+c_{2}^{*}{{a}^{*}}^{2}+...]\nonumber \\
&
\,\,\,\,\,\,\,\,\,\,\,\,\,\,\,\,=[1+2{{x}_{2}}a+3{{x}_{3}}{{a}^{2}}+...](-b{{a}^{2}})[1+ca+{{c}_{2}}{{a}^{2}}+...]\;.\label{star}\nonumber
\\
\end{align}
If we substitute ${{a}^{*}}$ from Eq.(\ref{starr}) on the first row of  Eq.(\ref{star}) and compare the result with the second row of this equation , then
one will arrive at
\begin{align}
& b={{b}^{*}} \nonumber\\
& c={{c}^{*}}\nonumber \\
& {{c}_{2}}=c_{2}^{*}+c{{x}_{2}}+x_{2}^{2}-{{x}_{3}}\nonumber \\
& .\nonumber \\
& .\nonumber \\
& .\nonumber \\
\end{align}
After this introductory subject we now deal with in details the McKeon $\it {et\; al.}$  approach.
\subsection{Renormalization group summation \label{AA}}
Here based on the McKeon {\it et al.} approach, it can be shown that by a full resumation on the QCD perturbative series the unphysical parameter, $\mu$,
would be disappeared and  the final result for any QCD observable, as it is expected, is independent of the unphysical parameter. One  of the feature of
McKeon {\it et al.} approach with respect to the CORGI approach is  based  on this fact that to achive this approach,   it does not need to use the self
consistency principle  and all required considerations to construct  the McKeon {\it et
al.} approach totally back to the RGE for QCD $\beta$-function. In CORGI approach one  needs to employ the self consistency principle not only for the
scheme parameters but also for  the scale parameter $\mu$. Therefore the scale and scheme parameters  are used simultaneously during the optimization
process
in CORGI approach  while in the McKeon
{\it et al.} approach,  a separation between the renormalization scale parameter in one hand side and scheme parameters on the other hand side
are realized perfectly.

Now we do a brief review on the McKeon {\it et al.} approach to show how the  unphysical parameters would be removed in perturbative series for any QCD
observable, just using RGE of QCD-$\beta$ function. For this purpose, by substituting Eq.(\ref{rn}) in Eq.(\ref{R}) the following result would be arrived
\cite{mac1}:
\be
R={{R}_{pert}}=\sum\limits_{n=0}^{\infty }{{{r}_{n}}{{a}^{n+1}}=\sum\limits_{n=0}^{\infty
}{\sum\limits_{m=0}^{n}{{{T}_{n,m}}{{L}^{m}}{{a}^{n+1}}}}}\;. \label{rp}
\ee
To satisfy the condition $r_0$=1 in Eq.(\ref{R}), it is required that $T_{0,0}=1$. From Eq.(\ref{rp}), the following expression should be assigned to
$r_{n}$:
\be
{{r}_{n}}=\sum\limits_{m}{{T}_{n,m}}{{L}^{m}}\;.
\ee
If one set $m=0$ then
\be
{{r}_{n}}={{T}_{n}}\Rightarrow R=\sum\limits_{n=0}^{\infty }{{{T}_{n}}{{a}^{n+1}}}\;, \label{rrtt}
\ee
which is similar to Eq.(\ref{R}).

Now a new grouping, denoted by $A_{n}$  is introduced \cite{mac1}:
\be
{{A}_{n}}=\sum\limits_{m=0}^{\infty }{{{T}_{n+m,n}}{{a}^{n+m+1}}}\;. \label{Aan}
\ee
This makes a possibility to sum the contribution to $R$, considering the RGE.
Therefore  using  Eq.(\ref{Aan}) in above, the $R$ in Eq.(\ref{rp}) would have the following presentation:
\be
R={{R}_{A}}=\sum\limits_{n=0}^{\infty }{{{A}_{n}}(a){{L}^{n}}}\;.\label{RA}
\ee
Substituting Eq.(\ref{RA}) in Renormalization Group Equation,  Eq.(\ref{RG}), it can be shown:
\be
\sum\limits_{n=0}^{\infty }{(bn{{A}_{n}}(a){{L}^{n-1}}+\beta (a){{{{A}'}}_{n}}(a){{L}^{n}})=0}\;.
\ee
Using this equation and rearranging the order of sums,  $A_{n}$ can be written as:
\be
{{A}_{n}}(a)=-\frac{\beta (a)}{nb}\frac{d}{da}{{A}_{n-1}}(a)\;.
\ee
Considering QCD  $\beta$-function, Eq.(\ref{beta}), one will arrive at:
\be
{{A}_{n}}(a(\ln \frac{\mu }{\Lambda }))=-\frac{1}{n}\frac{d}{d\ln (\frac{\mu }{\Lambda })}{{A}_{n-1}}(a(\ln
\frac{\mu }{\Lambda }))\label{An}\;,
\ee
where $\Lambda$ is related to be boundary condition on Eq.(\ref{beta}) such that:
\be
\ln (\frac{\mu }{\Lambda })=\int_{0}^{a}{\frac{dx}{\beta (x)}+\int_{0}^{\infty
}{\frac{dx}{b{{x}^{2}}(1+cx)}}}\label{49}\;.
\ee
Defining  $\eta =\ln \frac{\mu }{\Lambda }$ and using  the recurrence  relation,  Eq.(\ref{An}), the following result would be obtained:
\be
{{A}_{n}}(a(\eta ))=\frac{-1}{bn}\frac{d}{d\eta }{{A}_{n-1}}(a(\eta ))=\frac{1}{n!}{{(-\frac{1}{b}\frac{d}{d\eta
})}^{n}}{{A}_{0}}(a(\eta ))\label{An1}
\ee
Substituting Eq.(\ref{An1}) into Eq.(\ref{RA}) will lead to
\be
{{R}_{A}}=\sum\limits_{n=0}^{\infty }{\frac{1}{n!}{{(-\frac{L}{b})}^{n}}\frac{{{d}^{n}}}{d{{\eta
}^{n}}}{{A}_{0}}(a(\eta ))={{A}_{0}}(a(\eta -\frac{L}{b}))}\;.\label{ra}
\ee
To prove the above relation, it is easily to show:
\ba
R({A}) &=& \sum\limits_{n = 0}^\infty  {\frac{1}{{n!}}{{(-\frac{L}{b})}^n}\frac{{{d^n}}}{{d{\eta ^n}}}} {A_0}(a(\eta ))\nonumber\\
&&= \exp ( (-\frac{L}{b})\frac{d}{{d\eta }}){A_0}(a(\eta ))\;.\label{exp}\ea
On the other hand, using the Taylor expansion for the right hand side of Eq.(\ref{ra}), one can write
\ba &&{A_0}(a(\ln \frac{\mu }{\Lambda } - L)) = {A_0}(a(\ln \frac{\mu }{\Lambda })) - L{A'_0}(a(\ln \frac{\mu }{\Lambda })) \nonumber\\
&&+\frac{{{L^2}}}{{2!}}{A''_0}(a(\ln \frac{\mu }{\Lambda }))+ ...\;\;. \;,\ea
which is equivalent to the right hand side of Eq.(\ref{exp}), considering the expansion of exponential term.
Consequently, based on  the definitions of $L$ and $\eta$ parameters, one can see that \cite{mac2}:
\be
{{R}_{A}}={{A}_{0}}(a(\ln \frac{Q}{\Lambda }))\;.\label{A0}
\ee
This equation shows that all dependence  of $R$ on  $\mu$ scale  has been cancelled. This is a pleasant result since $\mu$
is an unphysical parameter and it is expected to be removed it by  doing a full resumation on the QCD perturbative series.
\subsection{Renormalization scheme dependence \label{B}}
In previous section, it has been shown that the final result for pertutbartive series of any QCD observable is independent of  renormalization scale,
$\mu$, as unphyscical parameter. Here a brief review on the McKeon  {\it et al.} approach is done to indicate how the perturbative series of QCD
observable would be scheme independent and therefore the final result for the QCD perturbative series is reliable and independent of the scale and scheme
parameters.

Let us to start the subject by reminding that the first two $b$ and $c$ parameters in QCD $\beta$- function are independent of renormalization scheme
while
the expansion parameters $c_{i}(i\geq2)$ are renormalization scheme dependent. It is now shown explicitly how
$R_{A}$ in Eq.(\ref{RA}) depends on $c_{i}$ parameters. Since $R_{A}$ is arising out from the sum over all perturbative terms, it should be  independent
of
any choice for renormalization scheme. Therefore one can write \cite{steven}:
\be
(\frac{\partial }{\partial {{c}_{i}}}+{{\beta }_{i}}(a)\frac{\partial }{\partial a}){{R}_{A}}=0\label{53}\;.
\ee
Here ${{\beta }_{i}}(a)$ is defined by
\be
\frac{\partial a}{\partial {{c}_{i}}}={{\beta }_{i}}(a)=-\widehat{\beta
}(a)\int_{0}^{a}{\frac{{{x}^{i+2}}}{{{(\widehat{\beta }(x))}^{2}}}dx}\label{betai}\;,
\ee
which indicates how the coupling constant depends on the   $c_{2},c_{3},...$ scheme parameters where $\widehat{\beta }(a)=\beta (a)/b$.
The solution of this equation is as following \cite{steven}:
\ba
&&\frac{\partial a}{\partial {{c}_{i}}}=\frac{1}{i-1}{{a}^{i+1}}\left[ 1-\frac{(i-2)}{i}ca+\right.\nonumber\\
&&\left( \frac{(i-1)(i-2)}{i(i+1)}{{c}^{2}}-\frac{(i-3)}{(i+1)}{{c}_{2}}\right){{a}^{2}}+...
\left.\right]\label{betai1}\;.
\ea
By setting ${{T}_{n,0}}\equiv {{T}_{n}}$ in Eq.(\ref{rp}) and based on Eq.(\ref{53}) one can derive the following result:
\be
\sum\limits_{n=0}^{\infty }{{{a}^{n+1}}\frac{\partial {{T}_{n}}}{\partial {{c}_{i}}}+(n+1){{\beta
}_{i}}(a){{T}_{n}}{{a}^{n}}=0}\;.
\ee
This leads to a set of nested equations for $T_{n}$:
\begin{align}
& \frac{\partial {{T}_{0}}}{\partial {{c}_{i}}}=0\,\,\Rightarrow {{T}_{0}}={{\tau }_{0}}=1\;, \nonumber\\
& \frac{\partial {{T}_{1}}}{\partial {{c}_{i}}}=0\Rightarrow {{T}_{1}}={{\tau }_{1}}=Const. \nonumber\label{55}\\
\end{align}
Reminding that $T_{0,0}=T_0=1$.\\

For the $T_2$ coefficients one can derive
\be
\frac{\partial {{T}_{2}}}{\partial {{c}_{2}}}+1=0\Rightarrow {{T}_{2}}=-{{c}_{2}}+{{\tau }_{2}}\label{56}\;,
\ee
where $\tau_2$ is again constant and an scheme invariant.\\

For the $T_3$ coefficient one leads to the two following set of equations:
\begin{align}
  & \frac{\partial {{T}_{3}}}{\partial {{c}_{2}}}+2{{\tau }_{1}}=0\,\,\,\;,\nonumber \\
 & \frac{\partial {{T}_{3}}}{\partial {{c}_{3}}}+\frac{1}{2}=0\;. \nonumber \\
\end{align}
Simultaneous solution of these two differential equations will end to

\be
{{T}_{3}}=-2{{c}_{2}}{{\tau }_{1}}-\frac{1}{2}{{c}_{3}}+{{\tau }_{3}}\label{58}\;,
\ee
where again $\tau_3$ is constant of integration and scheme invariant.\\

For the $T_4$ coefficient, one will get:
\begin{align}
& \frac{\partial {{T}_{4}}}{\partial {{c}_{2}}}+\frac{1}{3}{{c}_{2}}+3{{T}_{2}}=0\;, \nonumber\\
 & \frac{\partial {{T}_{4}}}{\partial {{c}_{3}}}+{{T}_{1}}-\frac{c}{6}=0\nonumber\;, \\
 & \frac{\partial {{T}_{4}}}{\partial {{c}_{4}}}+\frac{1}{3}=0 \nonumber\label{59}\;.\\
\end{align}
Solving the differential equations in Eq.(\ref{59}) compatibly with each other, give us the following result:
\be
{{T}_{4}}=-\frac{1}{3}{{c}_{4}}-{{c}_{3}}({{\tau }_{1}}-\frac{c}{6})+\frac{4}{3}c_{2}^{2}-3{{c}_{2}}{{\tau
}_{2}}+{{\tau }_{4}}\;,\label{60}
\ee
where once again $\tau_{4}$ is constant of integration and $RS$ invariant.\\

Following the procedure, introduced in \cite{mac1}  two specific choices of renormalization scheme are considered. In the first
scheme, the $c_{i}$ are selected so that $T_{n}=0\;(n\geq2)$. By this choice one then find from Eqs.Eqs.(\ref{55},\ref{56},\ref{58},\ref{60})
the following results:
\begin{align}
  & {{c}_{2}}={{\tau }_{2}}\;,\,\,\, \nonumber\\
 & \,{{c}_{3}}=2(-2{{c}_{2}}{{\tau }_{1}}+{{\tau }_{3}})\;,\nonumber \\
 & {{c}_{4}}=-\frac{3}{2}{{c}_{3}}(-\frac{c}{3}+2{{\tau }_{1}})+4c_{2}^{2}-9{{c}_{2}}{{\tau }_{2}}+3{{\tau
 }_{4}}\;. \label{61}
\end{align}
In this case, the expansion series in  Eq.(\ref{rp}) or equivalently Eq.(\ref{A0})  contains  just two terms \cite{mac1}:
\be
{{R}_{(1)}}={{a}_{(1)}}+{{\tau }_{1}}a_{(1)}^{2}(\ln \frac{Q}{\Lambda })\;,\label{62}
\ee
where the coupling ${{a}_{(1)}}$ in addition to universal parameters $b$ and $c$, depends also  on the $c_2$, $c_3$ and other scheme parameters. It should
be noted that since Eq.(\ref{A0}) is independent of renormalization scale $\mu$, the finite expansion in Eq.(\ref{62}) has been written in terms of
physical energy scale $Q$. As can be seen the result for the observable $R$ is scheme independent since $\tau_1$ is RS invariant and coupling constant
$a_{(1)}$  is independent of renormalization scale.\\

In the second case, the  scheme parameters are used such that $c_{i}=0\;(i\geq2)$ which is corresponding to the  't Hooft scheme \cite{thooft}. With this
choice, considering again Eqs.(\ref{55},\ref{56},\ref{58},\ref{60}) will lead to  $T_{n}=\tau_{n}$. In
this case Eq.(\ref{rp}) contains the infinite series:
\ba
&&{{R}_{(2)}}=\sum\limits_{n=0}^{\infty }{{{\tau }_{n}}a_{(2)}^{n+1}(\ln \frac{Q}{\Lambda })}={{a}_{(2)}}(\ln
\frac{Q}{\Lambda })+{{\tau }_{1}}a_{(2)}^{2}(\ln \frac{Q}{\Lambda })\nonumber\\
&&+{{\tau }_{2}}a_{(2)}^{3}(\ln \frac{Q}{\Lambda })+...\;\;.\label{63}
\ea
Like the expansion in Eq.(\ref{62}) and with the similar reason, just the physical energy scale $Q$ is appearing in series expansion of Eq.(\ref{62})
while
$\tau_1$,$\tau_2$ and etc. are RS invariant.

The $a_{(2)}$ coupling in Eq.(\ref{63}) is obtained from the solution of QCD-$\beta$ function while just two universal parameters $b$ ad $c$ are kept
there.
Therefore it can be expressed in terms of  the Lambert-$W$ function . For details, it is required to back to Eq.(\ref{beta}) while the general solution
reduces to:
\ba
&&\frac{1}{a}+c\ln (\frac{ca}{1+ca})=\tau
-\int{\left(\frac{1}{-{{a}^{2}}(1+ca+{{c}_{2}}{{a}^{2}}+...)}\right.}\nonumber\\
&&{\left.+\frac{1}{{{a}^{2}}(1+ca)}\right)da}\label{int}\;,
\ea
where $\tau$ is RS parameter, defined by $\tau =b\ln (\frac{Q}{\Lambda })$.
By setting $c_{n}=0$ in Eq.(\ref{int}), one will get:
\be
\frac{1}{{{a}_{0}}}+c\ln (\frac{c{{a}_{0}}}{1+c{{a}_{0}}})=b\ln (\frac{Q}{{{\Lambda }_{R}}}) \;.
\ee
To solve this equation the $W(Q)$ function is defined  such that:
\be
1+W(Q)=-\frac{1}{c{{a}_{0}}}\label{w1} \;.
\ee
Here $a_{0}=a_{0}(Q)$. Finally one  can write:
\be
-\frac{1}{c{{a}_{0}}}-\ln (\frac{c{{a}_{0}}}{1+c{{a}_{0}}})=\ln {{(\frac{Q}{{{\Lambda
}_{R}}})}^{-b/c}}\Rightarrow W{{e}^{W}}=-\frac{1}{e}{{(\frac{Q}{{{\Lambda }_{R}}})}^{-b/c}}\label{w}\;.
\ee
If $z(Q)=-\frac{1}{e}{{(\frac{Q}{{{\Lambda }_{R}}})}^{-b/c}}$ is introduced then Eq.(\ref{w}) will lead to:
\be
W(z){{e}^{W(z)}}=z\;. \ee
The solution of this equation is called Lambert W-function \cite{r4c,r5} then  from Eq.(\ref{w1}) the following result can be obtained:
\be
a_{(2)}\equiv {{a}_{0}(Q^2)}=-\frac{1}{c[1+{{W}_{-1}}(z(Q))]}\label{lam}\;.
\ee

However two different cases  for the McKeon {\it et al.} approach have been introduced but it can be shown that the expansions for $R_{(1)}$ and
$R_{(2)}$ given by Eq.(\ref{62}) and Eq.(\ref{63}) are equivalent to each other.  For this purpose
two different couplings $a_{c}$ and $a_{d}$ are considered which evaluated by  different renormalization schemes associated with
parameters $c_{i}$ and $d_{i}$  respectively \cite{mac1}:
\be
{{a}_{c}}={{a}_{d}}+{{\lambda }_{2}}({{c}_{i}},{{d}_{i}})a_{d}^{2}+{{\lambda
}_{3}}({{c}_{i}},{{d}_{i}})a_{d}^{3}+...\;.
\ee
The couplings ${{a}_{c}}$ and ${{a}_{d}}$ are being satisfied, each one, by their related QCD $\beta$-functions such that
\begin{align}
  & {{\beta }_{c}}({{a}_{c}})=-ba_{c}^{2}(1+c{{a}_{c}}+{{c}_{2}}a_{c}^{2}+...)\;, \nonumber\\
 & {{\beta }_{d}}({{a}_{d}})=-ba_{d}^{2}(1+c{{a}_{d}}+{{d}_{2}}a_{d}^{2}+...)\;.\nonumber \\
\end{align}
Since $\frac{d{{a}_{c}}}{d{{d}_{i}}}=0$  then it can be written
\be
(\frac{\partial }{\partial {{d}_{j}}}+{{\beta }_{j}}({{a}_{d}})\frac{\partial }{\partial
{{a}_{d}}})\sum\limits_{n=1}^{\infty }{{{\lambda }_{n}}({{c}_{i}},{{d}_{i}})a_{d}^{n}=0}\;.
\ee
Using the boundary condition ${{\lambda }_{n}}({{c}_{i}},{{c}_{i}})=0$, a set of differential
equations can be obtained for $\lambda_{n}$ whose solutions lead to the following relation for couplings:
\begin{align}
& {{a}_{c}}={{a}_{d}}-({{d}_{2}}-{{c}_{2}})a_{d}^{3}-\frac{1}{2}({{d}_{3}}-{{c}_{3}})a_{d}^{4}\nonumber \\
&+\left[ -\frac{1}{6}(d_{2}^{2}-c_{2}^{2})+\frac{3}{2}{{({{d}_{2}}-{{c}_{2}})}^{2}}\right.\nonumber \\
&+\frac{c}{6}({{d}_{3}}-{{c}_{3}})-\frac{1}{3}({{d}_{4}}-{{c}_{4}})\left.
\right]a_{d}^{5}+...\;.\nonumber\label{67} \\
\end{align}
Substituting this equation in Eq.(\ref{62}) and using 't Hooft scheme in which  $d_{i}=0\;(i\geq2)$ then the expansion for  $R_{(1)}$ reads to
\be
{{R}_{(1)}}=a_d+{{c}_{2}}{{a_d}^{3}}+(1/2){{c}_{3}}{{a_d}^{4}}+{{\tau
}_{1}}{{(a_d+{{c}_{2}}{{a_d}^{3}}+(1/2){{c}_{3}}{{a_d}^{4}})}^{2}}\;.\label{ad}
\ee
Evaluating $c_{i}$ from Eq.(\ref{61}) in terms of $\tau_i$s,
and substituting them in above equation, after proper  rearranging the desired series expansion, one  will arrive at \cite{mac1}:
\be
{{R}_{(1)}}=a_{(2)}+{{\tau }_{1}}{{a_{(2)}}^{2}}+{{\tau }_{2}}{{a_{(2)}}^{3}}+{{\tau }_{3}}{{a_{(2)}}^{4}}+...
\ee
which is complectly corresponding to series expansion for ${{R}_{(2)}}$ in Eq.(\ref{63}). Remembering that the $a_d$ coupling in Eq.(\ref{ad}) is
evaluated while the scheme parameters $d_{i}=0\;(i\geq2)$. Therefore it can be shown by $a_{(2)}$ as the Lambert-W  function and this gives us a full
agreement between Eq.(\ref{62}) and Eq.(\ref{63}).
 \section{Considering Some observable in Mckeon  and {{\it at al}} and CORGI approaches\label{exam1}}
After achieving the required mathematical framework for  the  McKeon   {\it {et al.}} approach, we  now examine it, considering two QCD
observable. We do their numerical calculations and compare them in further section with the results which will be obtained from CORGI approach.
At the first we employ the McKeon   {\it {et al.}} approach to obtain  the ratio of cross sections for  the electron-
positron annihilation into hadrons at the center of mass energy  $\sqrt s={31.6}\; GeV$.

\subsection{Electron-positron annihilation in Mckeon  and {{\it at al}} approach \label{exam1A} }
The expansion up to fourth order for $R_{e^+e^-}$ as the ratio for electron-positron annihilation into hadrons with ${{{N}_{f}}=5}$
can be written as \cite{44,45}:
\be
R_{{{e}^{+}}{{e}^{-}}}(s)=\frac{11}{3}[1+{{a}_{s}}+1.40902a_{s}^{2}-12.80a_{s}^{3}-80.434a_{s}^{4}+...]
\ee
According to the notation of  Mckeon   {{\it at al}} approach, one can write
\be
{{\text{T}}_{0}}\text{=1}\,\,,\,{{\text{T}}_{1}}\text{=1}\text{.409}\,\,\text{,}\,\,{{\text{T}}_{2}}
\text{=-12}\text{.80}\,\,\text{,}\,\,{{\text{T}}_{3}}\text{=-80}\text{.434}
\ee
Using Eqs.(\ref{55},\ref{56},\ref{58}) the numerical values for the required  RS invariants are given by:
\begin{align}
&{{\tau }_{1}}\text{=}{{\text{T}}_{1}}\text{=1}\text{.409}\,\,\,\text{,}\,\,\,\,
{{\tau }_{2}}\text{=}{{\text{T}}_{2}}\text{+}{{\text{c}}_{2}}\,=-11.3252\,\,\,\nonumber \\
&{{\tau }_{3}}\text{=}{{\text{T}}_{3}}\text{+}\frac{1}{2}{{\text{c}}_{3}}\text{+2}{{\text{c}}_{2}}{{\tau
}_{1}}\text{=-71}\text{.3600}\nonumber \\
\end{align}
We take the number of active quark flavour $N_{f}=5$ at  colliding energy  $\sqrt s={31.6}\; GeV$ and consider the QCD cut off value in the  $\overline{MS}$ scheme as
{{{}$\Lambda_{\overline{MS}}=419^{+222}_{-168}\; MeV$ which is raised from empirical value for ${R}_{{{e}^{+}}{{e}^{-}}}$ \cite{pmc12}. Final numerical result for the ratio of cross section in the Mckeon {{\it at al}} approach up to fourth order would be}}
\be
{{}\frac{3}{11}}{{R}_{{{e}^{+}}{{e}^{-}}}}=1+a+{{\tau }_{1}}{{a}^{2}}+{{\tau }_{2}}{{a}^{3}}+{{\tau }_{3}}{{a}^{4}}=1.0556^{+0.006}_{-0.006}\;,\label{ratio}
\ee
which is in good agreement with its available experimental value ${{R}_{{{e}^{+}}{{e}^{-}}}}=1.0527^{+0.005}_{-0.005}$  \cite{R-exp}. Reminding that
the coupling constant $a$ in Eq.(\ref{ratio}) is given in terms of Lambert-W function ( see Eq.(\ref{lam})).

There is a possibility to get more precise numerical value for ${{R}_{{{e}^{+}}{{e}^{-}}}}$ if we do  the calculations at first in Euclidean space and
then
by contour improved back the result to Minkowski space. This is because  the perturbative coefficients for the concerned observable have been computed
more
precisely in Euclidean space. In this regard we need to a relation between the ratio for $e^+e^-$ annihilation and Adler D-function  in Euclidean
space which can be found in \cite{Adler:1974gd}:
\be
\label{AdlerDispersion}
D(Q^2) = Q^2 \int_{4 m_\pi^2}^\infty \frac{R_{e^+e^-}(\cal{E})}{({\cal{E}}+Q^2)^2} {\rm d}{\cal{E}} \;.
\ee %$\pazocal{S}$
The Adler $D$-function would have the following expansion:
\be
{D}(Q^2)=a(1+{\sum_{n>0}}{d_n}{a}^{n}) \label{adlex}
\ee
where $d_i$  coefficients are known in \cite{44,45}. Nevertheless if one is interesting, can do some computations which finally yields $di$~s coefficients
in terms of $r_1$s ones. Similar computations but for Higgs decay width to gluon-gluon  in  details is done in Subsec.{\ref{exam2B}. Doing the same
calculations but for $R_{e^+e^-}$  will lead to:
\begin{align}\label{di-Adler}
& {{d}_{1}}={{r}_{1}}\,\,\,,\,\,\,\,\,{{d}_{2}}={{r}_{2}}+\frac{1}{3}\beta _{0}^{2}{{\pi }^{2}}\nonumber\;, \\
& {{d}_{3}}={{r}_{3}}+{{\pi }^{2}}{{r}_{1}}\beta _{0}^{2}+\frac{5}{6}{{\beta }_{0}}{{\pi }^{2}}{{\beta
}_{1}}\;.\nonumber \\
\end{align}
From Eqs.~(\ref{55},\ref{56}, \ref{58}) one can obtain $\tau_i$ as it
follows where in these equations  $T_i$ are replaced by $d_i$. Therefore one can write:
\be
{{\tau }_{1}}\text{=}{{\text{d}}_{1}}\,\,\,\,\,,\,\,\,\,\,\,{{\tau
}_{2}}\text{=}{{\text{d}}_{2}}\text{+}{{\text{c}}_{2}}\,\,\,\,\,\text{,}\,\,\,\,\,\,{{\tau
}_{3}}\text{=}{{\text{d}}_{3}}\text{+}\frac{1}{2}{{\text{c}}_{3}}\text{+2}{{\text{c}}_{2}}{{\text{d}}_{1}}\;.
\ee
To back the result to Minkowski space it is required to use  the analytic continuation then the observable $R$ can be written as \cite{mt}
\be  R({s}) = \frac{1}{{2\pi }}\int\limits_{ - \pi }^\pi  {W(\theta ) D({s}{e^{i\theta
}})d\theta }\;. \label{rrr} \ee
Here $W(\theta )$ is the weight function which  is taken  to be 1 for the ${R_{{e^ + }{e^ - }}}$. {{} Contour improved numerical result ,
taking ${\Lambda _{\overline {MS} }} = 419^{+222}_{-168}\; MeV$ at the center of mass energy $\sqrt s  = 31.6\,\,\,GeV$ would be}:
$\frac{3}{11}{R_{{e^ + }{e^ - }}} = 1.0547^{+0.005}_{-0.005}$  which is, as it expected, in better agreement with the available
experimental
data $\frac{3}{11} R^{\rm exp}_{e^+e^-}(\sqrt{s} = 31.6 \text{ GeV}) = 1.0527_{ - 0.005}^{ + 0.005}\;$ \cite{R-exp}.

{{{} If we take ${\Lambda _{\overline {MS} }} = 210^{+14}_{-14}\; MeV$ which is corresponding to world average value $\alpha_{s}(M_Z)=0.1181$  \cite{PDG} then we get $\frac{3}{11}{R_{{e^ + }{e^ - }}} = 1.0471^{+0.0006}_{-0.0007}$. The related conventional value for this observable is $\frac{3}{11}{R_{{e^ + }{e^ - }}} = 1.0461^{+0.0015}_{-0.0008}$ \cite{pmc16}.

Since there are  experimental data for ${R_{{e^ + }{e^ - }}}$ up to center of mass energy 208 $\text{ GeV}$ \cite{L3-2006} or even more, we do as well the required calculations in the McKeon {{\it et al.}} approach, taking  ${\Lambda _{\overline {MS} }} = 210^{+14}_{-14}\; MeV$ at the energy scales 42.5 and 56.5  $\text{ GeV}$. What we obtain for $\frac{3}{11}{R_{{e^ + }{e^ - }}}$ at these energy scales are 1.0463 and 1.0441 respectively which are comparable with their experimental values, i.e. 1.0554 \cite{Mark-J} and 1.0745 \cite{venus}. It seems that by increasing  energy scales the experimental data becomes little by little far away  from the theoretical prediction as can also be seen from \cite{Grif-2008} (see  Fig.8.4 of this Ref.). This is the reason we suspend  to consider the $R_{{e^ + }{e^ - }}$ at the other high energy scales. A summery of numerical results is listed in Table \ref{qua1}.}}
\subsection{Electron-positron annihilation in CORGI approach\label{exam1B}}
Here we are going to do the same calculation for electron-positron annihilation into hadrons but in CORGI approach. The required information for this
approach can be found in \cite{cj-1,Maxwell-Mir2000,PRD2019-AM}. We just remind that in this approach, using the self consistency principle while
the scheme parameters as well as renormalization scale are taken into account, it is possible to obtain an expression for pertutrbative coefficients of
QCD
observable at high
order in terms of  coefficients at lower order. Then doing the resumation over the perterbative expansion but  at a specified order, the resummed result
for instant at next-to-leading order (NLO), or the resummed result at next to NLO (NNLO) and etc. is such that to be RS invariant while
the renormalization scale is disappeared as well.
Hence there is a possibility to reconstruct the perturbative series in terms of RS invariants while this new constructed series does not depend as
well on renormalization scale as an unphysical parameter. By this brief introductory we are now intending to present the numerical result
for the $R_{{e^-}{e^+}}$ ratio in the CORGI approach.

The full theoretical expression for the electron-positron annihilation into hadrons can be written as \cite{ep}:
\be
R(s)=N{{\sum\limits_{f}{Q_{f}^{2}\left( 1+\frac{3}{4}{{C}_{F}}\widetilde{R}(s) \right)+\left(
\sum\limits_{f}{{{Q}_{f}}} \right)}}^{2}}\overline{R}(s)\;,\label{full}
\ee
where $\widetilde{R}$ is the perturbative corrections to the parton model result and has an expansion
as it follows:
\be
\widetilde{R}(s)=a(1+\sum\limits_{n}{{{r}_{n}}}{{a}^{n}})\;.
\ee
In  Eq.({\ref{full}}) $\overline{R}$ refers to ``light-by ligh'' contribution  and due to factor $\sum\limits_{f}{{Q}_{f}}^{2}$ which is zero for light quark flavors, it would be eliminate from the calculations.

{{
In the Minkowski space, based on the  CORGI approach, the perturbative part of  intended observable  has the following presentation \cite{PRD2019-AM}:
\be
{\widetilde R}(s)=a_0+X_{2}a_{0}^3+X_{3}a_{0}^4+...\;.\label{RCO}
\ee
 The $X_2$ and $X_3$   are scheme invariant quantities and according to notation of McKeon   ${\it et\; al.}$ approach can be written in  terms of $T_i$
 coefficients (see Eq.(\ref{rrtt}). The explicit expressions for these quantities have been presented in Ref.\cite{Mine-Akrami} (see Eq.~(26) of this
 reference). Then we will arrive at the these numerical results:
\be
X_2=-15.0870\,\,\,\,\,,\,\,\,\,\,X_3=-16.6423\;.
\ee
By substituting  the above  numerical values in  Eq.(\ref{RCO}) while $a_{0}$ is coupling constant at two loop levels which is written in  terms of
Lambert-W function and then inserting Eq.(\ref{RCO}) in Eq.(\ref{full}),  the following result for $R_{{e^-}{e^+}}$ ratio  in Minkowski space would be
obtained: $R_{e^+e^-}(\sqrt{s}=31.6
GeV)=1.05440\pm0.006$ . As before we take into account ${\Lambda _{\overline {MS} }} = 419^{+222}_{-168}\; MeV$   and the energy scale   $\sqrt s  =
31.6\,\,\,GeV$}}.

If one decides to do the calculations at first  in preferable Euclidean space, then the connection between  Adler D-function in this space with the
$\widetilde R(s)$ observable in
Minkowski space which is like Eq.(\ref{AdlerDispersion}), is again needed. The  perturbative expansion  of Adler D-function in Eq.(\ref{adlex}) has the
following form in CORGI approach
\be
\widetilde{D}(s)={{a}_{0}}+{{X}_{2}}a_{0}^{3}+{{X}_{3}}a_{0}^{4}+...+{{X}_{n}}a_{0}^{n+1}\label{ddd}\;.
\ee
Here $X_{i}$s are RS invariants and can be written in term of $d_{i}$ coefficients \cite{PRD2019-AM}. The $X_2$ and $X_3$ numerical results  for the Adler
D-function are as following :
\be
X_2=-7.2775\,\,\,\,\,\,,\,\,\,\,\,X_3=39.9935\;.
\ee
Using the analytic
continuation, given by Eq.(\ref{rrr}) and substituting  Eq.(\ref{ddd}) into it,  the numerical value  for  the full expression ${R}({{s}})$ in Minkowski
space can be obtained.
Numerical result for the concerned observable at the NNLO approximation, based on the CORGI approach, arouse out from analytical continuation and
taking $\Lambda_{\overline{MS}} = 419^{+222}_{-168}$ MeV would be $R_{e^+e^-}(\sqrt{s}=31.6 GeV)=1.0523\pm0.005$. This result
is in better agreement with  reported experimental data:~$1.0527\pm0.005$ \cite{R-exp} than the result, attained before, from direct calculations in
Minkowski space.

{{{}Like  to what we did in previous approach, we do as well the calculations in CORGI approach for $\frac{3}{11}{R_{{e^ + }{e^ - }}}$ at the energy scales 42.5 and 56.5  $\text{ GeV}$, taking $\Lambda_{\overline{MS}}=210\pm 14$. What we obtain  are 1.0436 and 1.0425 respectively while their experimental values are 1.0554 \cite{Mark-J} and 1.0745 \cite{venus}. Similar calculations in CORGI approach at $\sqrt s={31.6}\; GeV$  but with $\Lambda_{\overline{MS}}=210\pm 14$ have also been done. The related numerical values are listed in Table \ref{qua1}.}}
\section{Higgs decay to gluon in McKeon  {\it et al.} approach \label{exam2}}
As a second observable in McKeon  {\it ea al.} approach we take into account the  Higgs decay width to a gluon-gluon pair ($H\to gg$).  This observable
will be of fundamental phenomenological importance and its dominant contribution is given by \cite{hig4}:
\be
\Gamma (H\to gg)=\frac{4{{G}_{F}}M_{H}^{3}}{9\sqrt{2}\pi }R({{M}_{H}})\;,\label{gamma}
\ee
where $G_{F}$ is the fermi constant and $M_{H}$ is higgs boson mass. The numerical and perturbative series expansion of $R({M}_{H})$ is given later on in Subsec.\ref{exam2B} and Sec.\ref{cor} respectively.

As before we first review how the scale renormalization will be removed when we do a resummation over all perturbative terms of this observable. The
procedure to do RG summation for $H\to gg$ in many senses is like the one for $R_{{e^+}{e^-}}$ ratio. Nonetheless it involves its own worth to do it again
for this observable.  According
to the notation of McKeon  {\it ea al.} approach, the expansion series of  $R({{M}_{H}})$ has the following representation:
\be
R({{M}_{H}})=a_{s}^{2}\sum\limits_{n=1}^{\infty
}{\sum\limits_{m=0}^{n}{{{T}_{n,m}}a_{s}^{n}{{L}^{m}}}}\;,\label{gamma1}
\ee
where ${{T}_{0,0}}=1$ and  $L=\ln \frac{\mu }{{{M}_{H}}}$. Eq.(\ref{gamma1}) depends on renormalization scale   $\mu$ as unphysical quantity and it is
possible to show, as for the perturbative expansion in Eq.(\ref{rp}), that all dependence on $\mu$ in $\Gamma (H\to gg)$ can be removed by doing a
summation over all
logarithmic terms. We should note that the series expansion in Eq.(\ref{gamma1}) has an extra factor $a_{s}^{2}$ with respect to Eq.(\ref{rp}).
In this regard, the other grouping is introduced:
\be
{{A}_{n}}(a)=\sum\limits_{m=0}^{\infty
}{{{T}_{n+m,n}}{{a}^{n+m+2}}\,\,\,\,\,(n=0,1,2,...)\;.}\label{83}
\ee
Now Eq.(\ref{gamma}) can be represented by:
\be
\Gamma (H\to gg)=\frac{4{{G}_{F}}M_{H}^{3}}{9\sqrt{2}\pi }\sum\limits_{n=0}^{\infty }{{{A}_{n}}(a){{L}^{n}}}\;.
\ee
The RG equation implies that
\be
(\mu \frac{\partial }{\partial \mu }+\beta (a)\frac{\partial }{\partial a})\Gamma (H\to gg)=0 \;,
\ee
which like before leads to
\be
{{A}_{n}}(a)=-\frac{\beta (a)}{n}\frac{d}{da}{{A}_{n-1}}(a)\label{A}\;. \ee
By introducing $\eta$  scale as it follows \cite{mac2}
\be
\eta =\int_{{{a}_{I}}}^{a(\eta )}{\frac{dx}{\beta (x)}\,\,\,\,\,,\,\,\,\,{{a}_{I}}=const.}
\ee
Then
\be
\beta (a)\frac{d}{da}=\frac{d}{d\eta}\;.\label{eta}
\ee
By substituting Eq.(\ref{eta}) into Eq.(\ref{A}) one will arrive at
\be
{{A}_{n}}(a)=-\frac{1}{n}\frac{d}{d\eta }{{A}_{n-1}}(a(\eta ))\;,
\ee
where $\Lambda$ is defined by Eq.(\ref{49}). Now one can show that $A_0$, $A_1$,. . . and finally $A_n$ can be represented in terms of $A_0$. Therefore we
can write
\begin{align}
  & {{A}_{1}}=-1\frac{d}{d\ln \frac{\mu }{\Lambda }}{{A}_{0}}\;,\nonumber \\
 & {{A}_{2}}=\frac{-1}{2}\frac{d}{d\ln \frac{\mu }{\Lambda }}{{A}_{1}}=\frac{-1}{2}\frac{d}{d\ln \frac{\mu
 }{\Lambda }}(\frac{-1}{1}\frac{d}{d\ln \frac{\mu }{\Lambda }}{{A}_{0}})\nonumber \\
 &=\frac{1}{2}\frac{{{d}^{2}}}{d{{\ln
 }^{2}}\frac{\mu }{\Lambda }}{{A}_{0}}\;,\nonumber \\
 & {{A}_{3}}=-\frac{1}{3}\frac{d}{d\ln \frac{\mu }{\Lambda }}{{A}_{2}}=-\frac{1}{3}\frac{d}{d\ln \frac{\mu
 }{\Lambda }}\frac{1}{1.2}\frac{{{d}^{2}}}{d{{\ln }^{2}}\frac{\mu }{\Lambda
 }}{{A}_{0}}\nonumber \\
 &=\frac{-1}{1.2.3}\frac{{{d}^{3}}}{d{{\ln }^{3}}\frac{\mu }{\Lambda }}{{A}_{0}}\;,\nonumber \\
 & .\nonumber \\
 & . \nonumber\\
 & . \nonumber\\
 & {{A}_{n}}=\frac{{{(-1)}^{n}}}{n!}\frac{{{d}^{n}}}{{{d}^{n}}\ln \frac{\mu }{\Lambda }}{{A}_{0}}\;.\nonumber \\
\end{align}
If one defines $\eta =\ln \frac{\mu }{\Lambda }$  then
\be
{{A}_{n}}(a)=\frac{1}{n!}{{(-\frac{d}{d\eta })}^{n}}{{A}_{0}}(a)\;,
\ee
and finally
\ba
&&R({{M}_{H}})=\sum\limits_{n=0}^{\infty }{\frac{1}{n!}{{(-L)}^{n}}\frac{{{d}^{n}}}{d{{\eta
}^{n}}}}{{A}_{0}}(a(\eta ))\nonumber\\
&&=\exp (-L\frac{d}{d\eta }){{A}_{0}}(a(\eta ))\;.\label{RMHn}
\ea
The series expansion of  exponential term in Eq.(\ref{RMHn}) makes the right hand side of this equation  like the  Taylor expansion of
${{A}_{0}}(a(\ln \frac{\mu }{\Lambda }-L))$. For this purpose, as before one can write

\ba
&&{{A}_{0}}(a(\ln \frac{\mu }{\Lambda }-L))={{A}_{0}}(a(\ln \frac{\mu }{\Lambda }))-L{{{A}'}_{0}}(a(\ln
\frac{\mu }{\Lambda }))\nonumber\\
&&+\frac{{{L}^{2}}}{2!}{{{A}''}_{0}}(a(\ln \frac{\mu }{\Lambda }))+...\;\;.
\ea
Consequently  the sum in Eq.(\ref{RMHn}) results in
\ba
&&R({{M}_{H}})={{A}_{0}}(a(\eta -L))={{A}_{0}}(a(\ln \frac{\mu }{\Lambda }-\ln \frac{\mu
}{{{M}_{H}}}))\nonumber\\
&&={{A}_{0}}(a(\frac{{{M}_{H}}}{\Lambda }))\;.\label{94}
\ea
As shown by  Eq.(\ref{94}), the renormalization scale $\mu$ for Higgs decay width to gluon-gluon has been removed by doing a summation over all its
pertutbative
series terms.\\

Finally using Eq.(\ref{83}) with $m=0$, Eq.(\ref{94}) would be appeared as:
\be
\Gamma (H\to gg)=\frac{4{{G}_{F}}M_{H}^{3}}{9\sqrt{2}\pi }{{a(\frac{{{M}_{H}}}{\Lambda })}^{2}}(1+\sum\limits_{n=1}^{\infty
}{{{T}_{n}}{{a(\frac{{{M}_{H}}}{\Lambda })}^{n}})}\;.\label{95}
\ee
\subsection{Renormalization Scheme dependence for $H\rightarrow gg$ \label{exam2A}}
Here we specifically investigate how the series expansion for Higgs decay width to gluon-gluon can be  appeared in which it becomes RS invariant.

As before the RS dependence of the expansion coefficients $T_{n}$ in Eq.(\ref{95}) can be founded by using the RG equation
\be
(\frac{\partial }{\partial {{c}_{i}}}+{{\beta }_{i}}(a)\frac{\partial }{\partial a})\Gamma (H\to gg)=0\;,\label{betac}
\ee
where ${{\beta }_{i}}(a)$ has been given by Eq.(\ref{betai}). Considering perturbative part of Higgs decay in Eq.(\ref{95}) as
\be
R({{M}_{H}})={{a}^{2}}+\sum\limits_{n=1}^{\infty }{{{T}_{n}}{{a}^{n+2}}}\label{97}\;,
\ee
and employing  Eq.(\ref{betac}) we then arrive at
\be
2a{{\beta }_{i}}(a)+\sum\limits_{n=1}^{\infty }{{{a}^{n+2}}\frac{\partial {{T}_{n}}}{\partial
{{c}_{i}}}+(n+2){{a}^{n+1}}{{T}_{n}}}=0\;.
\ee
This equation leads to a sequence of  equations for $T_{n}$ order by order in terms of $a$ that their solutions are
\begin{align}
 & \frac{\partial {{T}_{1}}}{\partial {{c}_{i}}}=0\,\,\,\Rightarrow {{T}_{1}}={{\lambda }_{1}}\;, \nonumber\\
 & \frac{\partial {{T}_{2}}}{\partial {{c}_{2}}}+2=0\,\,\Rightarrow {{T}_{2}}=-2{{c}_{2}}+{{\lambda
 }_{2}}\nonumber\label{99}\;, \\
\end{align}
where ${{\lambda }_{1}}$ and ${{\lambda }_{2}}$ are constants of integrations and RS invariant. For $T_3$ coefficient, one can write
\begin{align}
& \frac{\partial {{T}_{3}}}{\partial {{c}_{2}}}+3{{T}_{1}}=0\;,\nonumber \\
 & \frac{\partial {{T}_{3}}}{\partial {{c}_{3}}}+1=0\;,\nonumber \\
\end{align}
which leads to
\be
{{T}_{3}}=-3{{\lambda }_{1}}{{c}_{2}}-{{c}_{3}}+{{\lambda }_{3}}\label{101}\;.
\ee
One can derive for $T_4$ coefficient in  similar manner the following differential equations:
\begin{align}
& \frac{\partial {{T}_{4}}}{\partial {{c}_{2}}}+\frac{2{{c}_{2}}}{3}+4{{T}_{2}}=0\;,\nonumber \\
& \frac{\partial {{T}_{4}}}{\partial {{c}_{3}}}-\frac{c}{3}+\frac{3}{2}{{T}_{1}}=0\;,\nonumber \\
& \frac{\partial {{T}_{4}}}{\partial {{c}_{4}}}+\frac{2}{3}=0\nonumber\;, \\
\end{align}
which leads to the final result:
\be
{{T}_{4}}=\frac{11}{3}c_{2}^{2}-4{{\lambda }_{2}}{{c}_{2}}+\frac{c}{3}{{c}_{3}}-\frac{3}{2}{{\lambda
}_{1}}{{c}_{3}}-\frac{2}{3}{{c}_{4}}+{{\lambda }_{4}}\;.\label{103}
\ee
This process can be followed to obtain the other expressions for higher order of $T_i$ coefficients.

Like for $T_1$ and $T_2$ the  $\lambda_{3}$ and $\lambda_{4}$ in coefficient of $T_3$ and $T_4$ are constant of integration and RS invariant that can be
determined once $T_{i}$ and $c_{i}$
have been evaluated in some mass independent RS.\\

Since the summation of concerned  perturbative series are scheme invariant, again two particular RS are of special interest. In the first scheme one can
set $c_{i}=0\; (i\geq2)$ and consequently
$T_{n}=\lambda_{n}$. In this case  Eq.(\ref{97}) is appearing as
\be
{{R}_{(2)}}({{M}_{H}})={{a_{(2)}}^{2}}+\sum\limits_{n=1}^{\infty }{{{\lambda }_{n}}{{a_{(2)}}^{n+2}}}\label{104}\;.
\ee
In the other case with setting $T_{i}=0\; (i\geq 2)$,  Eq.(\ref{97}) would involve just two terms
\be
{{R}_{(1)}}({{M}_{H}})={{a}^{2}}+{{\lambda }_{1}}{{a}^{3}}\;. \label{r11}
\ee
Reminding that  the coupling $a$  in Eq.(\ref{r11}) depends on the scheme parameters $c_i$ while  the coupling Eq.(\ref{104}) depends just on two $b$ and
$c$ universal scheme parameters that can be expressed in terms of the Lambert function $W(x)$ i.e
$x=W(x)e^{W(x)}$.

Again, considering the relation between the couplings at two different schemes,  one can show that two presentation of Eq.(\ref{r11}) and Eq.(\ref{104})
are equivalent such that
\be
{{R}_{(1)}}({{M}_{H}})={{R}_{(2)}}({{M}_{H}})\;. \ee
\subsection{Numerical value for Higgs decay to gluon\label{exam2B}}
In this section we do numerical investigation for the Higgs decay width to gluon-gluon in McKeon  {\it {et al.}} approach.
The Higgs boson decay to gluon-gluon is given by Eq.(\ref{gamma}) {{}where $R(M_{H})$ in this equation has a numerical  expansion} in conventional perturbative QCD
as it follows \cite{hig4}
\be
R({{M}_{H}})={{a}^{2}}(1+17.9167a+153.15787{{a}^{2}}+393.822{{a}^{3}}+...)\label{bast}\;.
\ee
According to the notation of
McKeon {\it {et al.}} approach, one can obtain from  Eq.(\ref{bast}) the following numerical values for the $T_i$ coefficients:
\be
{{\text{T}}_{1}}\text{=17}\text{.9167}\,\,\,\text{,}\,\,\,{{\text{T}}_{2}}\text{=153}\text{.1578788}\,\,\,\text{,}\,\,\,{{\text{T}}_{3}}\text{=
393}\text{.822046}\;.
\ee
From Eqs.(\ref{99},\ref{101}) the following numerical values for the $\lambda_i$ as RS invariants would be obtained:
\begin{align}
& {{\lambda }_{1}}\text{=}{{\text{T}}_{1}}\text{=17}\text{.9167}\nonumber \\
& {{\lambda }_{2}}\text{=}{{\text{T}}_{2}}\text{+2}{{\text{c}}_{2}}\text{=156}\text{.1074561}\nonumber \\
& {{\lambda }_{3}}\text{=}{{\text{T}}_{3}}\text{+}{{\text{c}}_{3}}\text{+3}{{\text{c}}_{2}}{{\lambda
}_{1}}=\text{482}\text{.927999}\nonumber \\
\end{align}
Now Eq.(\ref{104}) which has been obtained, using 't Hooft scheme,  becomes
\be
R({{M}_{H}})={{a_{(2)}}^{2}}+{{\lambda }_{1}}{{a_{(2)}}^{3}}+{{\lambda }_{2}}{{a_{(2)}}^{4}}+{{\lambda }_{3}}{{a_{(2)}}^{5}}+...\;.\label{rrhh}
\ee
In this equation, as illustrated before,  coupling constant $a_{(2)}$ can be written in terms of Lambert-W function which is  depends on physical energy
scale Q and $\lambda_i$ are specified scheme invariants. To do numerical calculation for the decay width $H\rightarrow gg$ in the McKeon  ${\it et\; al.}$
approach
we consider again $G_{F} =
1.16638\times10^{-5}\;GeV^{-2}$ and the Higgs mass $M_{H} = 126\pm4 GeV$ while we take the number of active quark flavor  $N_{f}=5$ and QCD cut off
$\Lambda_{\overline{MS}}=210\pm14$ MeV . What we achieve for the concerned decay width is: $\Gamma(H\rightarrow
gg) = 0.383\pm 0.01\; MeV$ which can be
compared to the result of the conventional value: $0.349\pm0.05$ MeV {{{} \cite{27,28}}}.

As for $R_{{e^+},{e^-}}$ ratio it is possible to follow the numerical calculation in Euclidian space and then by contour improved back to Minkoski space
to
get more precise result.

Back to Eq.({\ref{rrhh}}) the related  Adler-D  function should have the following expansion

\be
D({{M}_{H}})={{a}^{2}}(1+\sum\limits_{n=1}^{\infty
}{{{d}_{n}}{{a}^{n}})=}{{a}^{2}}+{{d}_{1}}{{a}^{3}}+{{d}_{2}}{{a}^{4}}+{{d}_{3}}{{a}^{5}}+...
\;.
\ee

There are the following relations between that $d_i$ coefficients and the $T_i$  ones in Eq.(\ref{97}) where we deal with about them in next section:
\begin{align}
  & {{d}_{1}}={{T}_{1}}\,\,\,\,\,,\,\,\,\,\,\,{{d}_{2}}={{T}_{2}}+\beta _{0}^{2}{{\pi }^{2}}\;, \\
 & {{d}_{3}}={{T}_{3}}+2{{\pi }^{2}}{{T}_{1}}\beta _{0}^{2}+\frac{7}{3}\beta {{\pi }^{2}}{{\beta }_{1}}\;. \\
\end{align}
Now  using Eqs. (\ref{99},\ref{101}) one can obtain $\lambda_i$ invariants as it follows:
\be
{{\lambda }_{1}}\text{=}{{\text{T}}_{1}}\,\,\,\,\,\,,\,\,\,\,\,\,\,{{\lambda
}_{2}}\text{=}{{\text{d}}_{2}}\text{+2}{{\text{c}}_{2}}\,\,\,\,\,\text{,}\,\,\,\,\,\,{{\lambda
}_{3}}\text{=}{{\text{d}}_{3}}\text{+}{{\text{c}}_{3}}\text{+3}{{\text{c}}_{2}}{{\lambda }_{1}}\;.
\ee
Finally the perturbative part in Minkowski space converts to
\be
{D}({{M}_{H}})={{a_{(2)}}^{2}}+{{\lambda }_{1}}{{a_{(2)}}^{3}}+{{\lambda }_{2}}{{a_{(2)}}^{4}}+{{\lambda }_{3}}{{a_{(2)}}^{5}}+...\;,
\ee
where in addition to $\lambda_i$  quantities as RS invariants, $a_2$ coupling that is written in terms of Lambert-W function, is independent of
renormalization scheme and depends just on physical energy scale $Q=M_H$. Now
the results for Higgs decay width to gluon-gluon in the McKeon   ${\it {et\; al.}}$ approach but in the Euclidian space is
accessible. Using the contour improved integration as in below we could get the decay width in the Minkowski
space:
\be
R({{M}_{H}})=\frac{1}{2\pi }\int\limits_{-\pi }^{\pi }{{D}({{M}_{H}}{{e}^{i\theta }})d\theta}\;.
\ee
Finally by taking $M(H)=126\pm 4\; GeV$ and $\Lambda _{QCD}^{{{n}_{f}}=5}=210\pm 14\;MeV$ the result would be $\Gamma(H\rightarrow
gg)=0.370\pm 0.01\;MeV$ which is compatible with the convention one: $0.349\pm 0.05\;MeV$. At the moment there is not any experimental data for Higgs
decay width to gluon-gluon. But it is expected the result, arising out from counter improved, shall be closer to experimental value. {{{} Similar calculations but with updated value for Higgs mass, $M_{H} = 125.18\pm0.16 GeV$ \cite{PDG} have also been done. The concerned numerical results have been brought in Table \ref{qua2}.}}
\section{Higgs decay to gluon in CORGI approach\label{cor}}
Here we review how to obtain the result for Higgs decay to gluon-gluon in CORGI approach. More details can be found in \cite{PRD2019-AM}.
Nevertheless in order the readers can follow the subject clearly, some main steps to arrive at CORGI approach  for $\Gamma (H\to gg)$ are given here.
Reminding that in conventional perturbative series, the Higgs decay width to gluon-gluon is given by
\ba
&&\Gamma (H\to gg)=\frac{4{{G}_{F}}M_{H}^{3}}{9\sqrt{2}\pi }\left[ a_{s}^{2}+{{r}_{1}}a_{s}^{3}\;,
\right.\nonumber\\
&&+{{r}_{2}}a_{s}^{4}+\left. {{r}_{3}}a_{s}^{5}+{\cal O} (a_{s}^{6}) \right]\label{Higs}
\ea
where $a=\frac{\alpha_{s}}{\pi}$ is coupling constant and coefficients $r_{i}$ are known in \cite{hig4}. As it is obvious the perturbative
part of Eq.(\ref{Higs}) is presented by:
\be
R={{a}^{2}}+{{r}_{1}}{{a}^{3}}+{{r}_{2}}{{a}^{4}}+{{r}_{3}}{{a}^{5}}+...\;.\label{113}
\ee
Self-consistency principle for this observable up to n-th order, yields us \cite{steven}
\be
\frac{\partial{R^{(n)}}}{\partial{(RS)}}={\cal O}(\alpha_{s}^{n+1})\;,
\ee
where $n=1,2,...,n$ and RS stands for  the scale parameter $\tau =b \ln \frac{\mu}{\Lambda}$ as well as the scheme parameters such as
${{c}_{2}},{{c}_{3}},{{c}_{4}},...$ \cite{Maxwell-Mir2000,PRD2019-AM}.
%where $n=1,2,...,n$ and RS stands for either the scale parameter $\tau =b \ln \frac{\mu}{\Lambda}$ and the scheme parameters such as
%${{c}_{2}},{{c}_{3}},{{c}_{4}},...$ as well as ${{r}_{2}},{{r}_{3}},...$ coefficients \cite{Max-Mir,mine-pmc}.
Considering  two terms in Eq.(\ref{113}) that is
\be
R={{a}^{2}}+{{r}_{1}}{{a}^{3}}\label{rto}\;,
\ee
 and using self-consistency principe with respect to $\tau$ variable, one then concludes
\be
\frac{\partial {{r}_{1}}}{\partial \tau }=2\Rightarrow {{r}_{1}}-2\tau ={{\rho }_{1}}\;.
\ee
Here $\rho_{1}$ is RS invariant  and independent of un-physical scale parameter $\tau$. Adding the
third term to Eq.(\ref{rto}) which contains the $r_{2}$ coefficient then it appears as:
\be
R(Q)={{a}^{2}}+{{r}_{1}}{{a}^{3}}+{{r}_{2}}{{a}^{4}}\;.
\ee
Then by derivation of  $R(Q)$ with  respect to the related RS parameters, one arrives at
\begin{align}
  & \frac{\partial {{r}_{2}}}{\partial {{c}_{2}}}=-2,\,\,\;, \nonumber\\
 & \,\frac{\partial {{r}_{2}}}{\partial {{r}_{1}}}=c+\frac{3}{2}{{r}_{_{1}}}\nonumber\label{117}\;, \\
\end{align}
where in deriving the first above  differential equation, the  Eq.(\ref{betai1}) has been used. Simultaneous solution of differential equations
(\ref{117})
will lead to:
\be
{{r}_{2}}=c{{r}_{1}}+\frac{3}{4}r_{1}^{2}-2{{c}_{2}}+{{X}_{2}}\;.\label{r118}
\ee
Here $X_{2}$ is the constant of integration and RS invariant.
If the forth term of the series expansion in Eq.(\ref{113}) is considered such that
\be
R(Q)={{a}^{2}}+{{r}_{1}}{{a}^{3}}+{{r}_{2}}{{a}^{4}}+{{r}_{3}}{{a}^{5}}\;,
\ee
then self consistency principle implies that
\begin{align}
  & \frac{\partial {{r}_{3}}}{\partial {{c}_{2}}}=-3{{r}_{1}}\;,\nonumber \\
 & \frac{\partial {{r}_{3}}}{\partial {{c}_{3}}}=-1\;, \nonumber\\
 & \frac{\partial {{r}_{3}}}{\partial {{r}_{1}}}=\frac{3}{2}c{{r}_{1}}+2{{r}_{2}}+{{c}_{2}}\;. \nonumber\\
\end{align}
Considering the above differential equations, the result for ${{r}_{3}}({{r}_{1}},{{c}_{2}},{{c}_{3}})$ is
\be
{{r}_{3}}({{r}_{1}},{{c}_{2}},{{c}_{3}})=\frac{r_{1}^{3}}{2}+\frac{7}{4}cr_{1}^{2}-3{{c}_{2}}{{r}_{1}}+2{{X}_{2}}{{r}_{1}}-{{c}_{3}}+{{X}_{3}}\;,\label{r121}
\ee
where $X_{3}$ is the constant of integration and RS invariant. Substituting Eqs.(\ref{r118},\ref{r121}) in
Eq.(\ref{113}) the result would be
\ba
&&R(Q)={{a}^{2}}+{{r}_{1}}{{a}^{3}}+(c{{r}_{1}}+\frac{3}{4}r_{1}^{2}-2{{c}_{2}}+{{X}_{2}}){{a}^{4}}\nonumber\\
&&+(\frac{r_{1}^{3}}{2}+\frac{7}{4}cr_{1}^{2}-3{{c}_{2}}{{r}_{1}}+2{{X}_{2}}{{r}_{1}}-{{c}_{3}}+{{X}_{3}}){{a}^{5}}+...\;.\nonumber\\
\ea
Doing  resummation for the individual NLO, NNLO and etc. contributions in above equation and  employing the 't Hooft scheme at any specified order such
that  ${{c}_{2}}={{c}_{3}}=...={{c}_{n}}=0$ and for convenience to set $r_{1}=0$ then one will arrive at the following result \cite{PRD2019-AM}
\be
R(Q)=a_{0}^{2}+{{X}_{2}}a_{0}^{4}+{{X}_{3}}a_{0}^{5}+...\;.\label{RCGG}
\ee
In this equation  $a_{0}$  and $X_{i}$ are scale and scheme invariant and $a_{0}$ can be expressed in term of Lambert-W
function.
{{
At first we present the numerical result for Higgs decay width to gluon-gluon directly in Minkowski space. For this propose we require initially to
calculate the numerical values for the $X_2$ and $X_3$ in Eq.(\ref{RCGG}) which can be done using  Eqs.(\ref{r118},\ref{r121}) where $r_i$ coefficients
have been calculated in [25].
The results for these RS invariant quantities are as fo;;owing:
\be
X_2= -107.2392\,\,\,\,\,,\,\,\,\,\,X_3= 3269.9755
\ee
By substituting the above results  in Eq.(\ref{RCGG}) and taking $M(H)=126\pm 4\; GeV$ and $\Lambda _{QCD}^{{{n}_{f}}=5}=210\pm10\; MeV$ the numerical
result for Higgs decay width would be:
$\Gamma(H\rightarrow
gg)=0.393\pm 0.025 \;MeV.$}}

Since the calculation can be done in Euclidean space as in the $R_{e^+e^-}$, it is needed to convert the
calculations from the Minkowski to Euclidean
space. For this propose at first one needs to a relation between coupling constants at different scales:
\be
a(Q)=a(\mu )(1+{{h}_{1}}a(\mu )+{{h}_{2}}a{{(\mu )}^{2}}+{{h}_{3}}a{{(\mu )}^{3}}+{{h}_{4}}a{{(\mu
)}^{4}}+...)\label{aQ}\;.
\ee
The following relations exist for the  $h_{i}$~s coefficients \cite{14}:
\begin{align}
& {{h}_{1}}=-{{\beta }_{0}}L\;,\,\,\,\,,\,\,\,\,\,\,{{h}_{2}}=\beta _{0}^{2}{{L}^{2}}-{{\beta
}_{1}}L\;,\,\,\,\,\nonumber \\
& \,{{h}_{3}}=-\beta _{0}^{3}{{L}^{3}}+\frac{5}{2}{{\beta }_{0}}{{\beta }_{1}}{{L}^{2}}-{{\beta }_{2}}L\;,\nonumber
\\
& {{h}_{4}}=\beta _{0}^{4}{{L}^{4}}-\frac{13}{3}\beta _{0}^{2}{{\beta }_{1}}{{L}^{3}}+3(\frac{1}{2}\beta
_{1}^{2}+{{\beta }_{0}}{{\beta }_{2}}){{L}^{2}}-{{\beta }_{3}}L\;. \nonumber\\
\end{align}
Here $L=\ln (\frac{s}{{{\mu }^{2}}})$ where $s$ is new scale  and $\beta_{i}$s are the coefficients of QCD $\beta$ function .
Substituting Eq.(\ref{aQ}) in Eq.(\ref{113}) will yield us the observable $R $ in terms of the $s$ variable. Using the
dispersion relation, Eq.(\ref{AdlerDispersion}), but for Higgs decay width to gluon-gluon, will yield us this observable in Euclidian space. To back the
result form the $\mu$ scale to the $Q
$ one, it
is needed to use again the displacement relation, Eq.(\ref{aQ}), in a proper state. Finally we arrive at
\ba
&&D(Q)=a{{(Q)}^{2}}+{{r}_{1}}a{{(Q)}^{3}}+({{r}_{2}}+\beta _{0}^{2}{{\pi }^{2}})a{{(Q)}^{4}}\nonumber \\
&&+(\frac{7}{3}{{\beta }_{0}}{{\beta }_{1}}{{\pi }^{2}}+{{r}_{3}}+2\beta _{0}^{2}{{\pi
}^{2}}{{r}_{1}})a{{(Q)}^{5}}+O{{(aQ)}^{6}}\label{DQ2}
\ea
By comparing  Eq.(\ref{113}) and Eq.(\ref{DQ2}) the following relations between $d_{i}$~s and $r_i$~s coefficients would be obtained:
\begin{align}\label{di-Adler}
& {{d}_{1}}={{r}_{1}}\,\,\,,\,\,\,\,\,{{d}_{2}}={{r}_{2}}+\beta _{0}^{2}{{\pi }^{2}}\nonumber \\
& {{d}_{3}}={{r}_{3}}+2{{\pi }^{2}}{{r}_{1}}\beta _{0}^{2}+\frac{7}{3}{{\beta }_{0}}{{\pi }^{2}}{{\beta
}_{1}}\nonumber \\
\end{align}
Using the counter improved integration as in below we could get back the decay width to the Minkowski
space:
\be
\Gamma (H\to gg)=\frac{4{{G}_{F}}M_{H}^{3}}{9\sqrt{2}\pi }\frac{1}{2\pi }\int_{0}^{2\pi
}{[a_{0}^{2}+{{X}_{2}}a_{0}^{4}+{{X}_{3}}a_{0}^{5}+...]d\theta }\;.
\ee
where
\be
X_2=472.8741\,\,\,\,\,,\,\,\,\,\,X_3=10096.8174
\ee
To compute the above numerical results, we need again to Eqs.(\ref{r118},\ref{r121}) while $r_i$ coefficients are replaced by $d_i$ ones, given by
Eq.(\ref{di-Adler}).
As before ${{a}_{0}}({{s}_{0}}{{e}^{i\theta }})$ is coupling constant in CORGI approach.
To do numerical calculation for the decay width $H\rightarrow gg$ in the CORGI approach, we take as before $G_{F} =
1.16638\times10^{-5}\; GeV^{-2}$ and the Higgs mass $M_{H} = 126\pm4\; GeV$ while we take $N_{f}=5$ and
$\Lambda_{\overline{MS}}=210\pm14\; MeV$ . What we
obtain for the concerned decay width would be:$\Gamma(H\rightarrow
gg) = 0.379\pm 0.03\; MeV$ which can be
compared with the conventional result : $0.349\pm0.05\; MeV$ \cite{27,28}. {{{}The numerical results, considering updated value for Higgs mass, $M_{H} = 125.18\pm0.16 GeV$ \cite{PDG}, have been added in Table \ref{qua2}}}.
\begin{table*}[ht]
%\begin{table}[ht]
\begin{center}
\begin{tabular}{|c|c|c|c|c|c|}
\hline $$&McKeon ${\it et\; al.}$ approach&CORGI apporch&Conventional pQCD&Experimental value&$\Lambda_{\overline{MS}}$\\
\hline $R_{e^{+}e^{-}}( \sqrt{s}= 31.6 GeV) $&$1.04711_{-0.0007}^{+0.0006}$&$1.04615_{-0.00005}^{+0.00003}$&$1.04617_{-0.00008}^{+0.00015}$&$1.0527_{-0.005}^{+0.005}$&$210\pm 14 MeV$\\
\hline $R_{e^{+}e^{-}}(\sqrt{s}= 42.5 GeV) $&$1.0463_{-0.0004}^{+0.0004}$&$1.0436_{-0.0005}^{+0.0004}$&$1.0437_{-0.0005}^{+0.0004}$&$1.0554\pm 0.2$&$210\pm 14 MeV$\\
\hline $R_{e^{+}e^{-}}( \sqrt{s}= 52.5 GeV) $&$1.0441_{-0.0004}^{+0.0005}$&$1.0425_{-0.0004}^{+0.00045}$&$1.0424_{-0.0004}^{+0.0004}$&$1.0745\pm 0.11$&$210\pm 14 MeV$\\
%\hline $R_{e^{+}e^{-}}(Q=91.1876 GeV)$&$1.0391_{-0.0003}^{+0.0003}$&$1.0387_{-0.0002}^{+0.0002}$&$1.03884$&--&$210\pm 14 MeV$\\
\hline
\end{tabular} \label{qua1}
\caption{Numerical results for $R_{e^{+}e^{-}}$ at $\sqrt{s}= 31.6, 42.5\; \text{and}\; 56.5\;GeV$, using two different approaches.}\label{qua1}
\end{center}
\end{table*}
\begin{table*}[ht]
%\begin{table}[ht]
\begin{center}
\begin{tabular}{|c|c|c|c|c|c|}
\hline $$&McKeon ${\it et\; al.}$ approach&CORGI apporch&Conventional pQCD&Experimental value&$\Lambda_{\overline{MS}}$\\
\hline $H\rightarrow gg(M_{H}=126\pm4)$&$0.370_{-0.01}^{+0.01}$&$0.379_{-0.02}^{+0.02}$&$0.349_{-0.007}^{+0.007}$&--&$210\pm14 MeV$\\
\hline $H\rightarrow gg(M_{H}=125.18\pm0.16)$&$0.369_{-0.0020}^{+0.0018}$&$0.378_{-0.0015}^{+0.0018}$&$0.349_{-0.007}^{+0.007}$&--&$210\pm14 MeV$\\
\hline
\end{tabular} \label{qua}
\caption{Numerical results for Higgs decay to gloun-gloun  at two different approaches.}\label{qua2}
\end{center}
\end{table*}
\section{conclusion}\label{con}
In this article, we  reviewed first the principles that the McKeon  ${\it et\; al.}$ approach has been based on  them \cite{mac1,mac2,mac3,mac4,mac5}.
In this approach two essential steps have been done. At first by defining a new summed  group, denoted by $A(n)$, it was tried by taking into account  the
RGE,  to find a
pattern which gives a recurrence relation between the $n^{th}$ and the $(n-1)^{th}$ order of $A(n)$. Finally the concerned quantity at $n^{th}$
order can be written in terms of the its  first order. In deriving this pattern the role of QCD $\beta$-function is very essential. Then by constructing
the  primary series of QCD observable in terms of $A(n)$ and  doing a resummation over
all perturbative terms, the renormalization scale as an unphysical quantity was disappeared and the final result for the QCD observable
depended just on physical energy scale $Q$ (see Eq.(\ref{A0})).

In the second step of this approach, RS dependence of  perturbative series coefficients were being investigated. Considering the
differential
equation for the coupling constant with respect to scheme parameters $c_i,(i\geqslant 2)$, based on the Eq.(\ref{betai}) one could rewrite the
perturbative
series of QCD observable in terms of scheme invariant quantities while the coupling constant depends just on the physical energy scale, written
in
terms of Lambert-W function (see Eq.(\ref{63})). It should be reminded that in this approach there is a possibility to render the RS invariant and scale
independence of series expansion such that the series contains just two terms. In this case the related coupling constant, however is scale independence,
but depends on the
scheme parameters $c_2$, $c_3$ and etc.  But  the series is overall independent of renornalization scale and scheme (see Eq.(\ref{62})). At the end of
subsection $\ref{B}$ it has been shown that these two series expansion are
equivalent.

In the second approach which was called CORGI, using the self consistency principle there would be a possibility to express the $r_i$ coefficient at
$n^{th}$ order in terms of  the coefficients at lower orders which finally include just the coefficient at first order, i.e.  $r_1$. During to employ this principle, not
only the derivative of
observable with respect to scheme parameters $c_i (i\geqslant 2)$  is considered  but also the  derivative with respect   to $r_1$ coefficient is
required to obtain
such that in  $\frac{\partial R^{(i)}}{\partial {(RS)}}$ derivative, the $RS^,s$ can be labeled by $(r_1,c_2,c_3, . . .)$ parameters \cite{steven}. The
parameter $r_1$ can be casted latter on in terms of $\tau=\ln \frac{\mu}{\Lambda}$ as scale parameter \cite{PRD2019-AM} . By simultaneous solution of
differential equations which have been obtained for  $r_i$, using self-consistency principle, with respect to  $(r_1,c_2,c_3, . . .)$ parameters, the
$r_i $ coefficients  can be presented as a function of theses parameters. Rewriting the initial perturbative series of QCD observable, Eq.(\ref{R}) in
terms of relations which have been attained for $r_i$ coefficients, a possibility would be  provided  to do a  resummation  over the parturbative series
at individual NLO approximation, later on
at NNLO approximation and so on. Then the desired perturbative series can be rewritten in terms of scheme invariants and coupling constant which just
depends on physical energy scale.

From this point of views the CORGI approach and the the McKeon  ${\it et\; al.}$ approach have been achieved the same goal. Both of them were able to
reconstruct the conventional perturbative series in terms of the scheme invariants and the coupling constant which does not depend on renormalization
scheme. But which one has more advantage with respect to the other one?. At first consideration,  one may assign the priority to the McKeon  and ${\it
et\;
al.}$
approach. In this approach the renormalization scale parameter and scheme parameters are  investigating separately, as we explained it in
introduction
section while in CORGI approach this separation, utilizing the self-consistency principle, has not occurred.

We reexamined these two approaches and employed them to do numerical computations for $R_{e^+e^-}$ ratio and the Higgs decay width to gluon-gluon. We
listed the numerical results, based on the contour improved integration, in Tables $\ref{qua1}$ and $\ref{qua2}$. Inspection the Table $\ref{qua1}$ indicates that the results of McKeon  ${\it et\; al.}$ approach are in better agreement with the available data.

In addition to these two approaches, there is another approach which is called principle of maximum comfortability (PMC).  We explained about this
approach in introduction section. Based on this approach there is a possibility to absorb the non-conformal parts of the  perturbative series coefficients
into the renormailized coupling constant and finally to attain a perturbation series in terms of just conformal parts while the renormalization scales are
also fixed at any specified order. It is possible to convert the  McKeon   ${\it et\; al.}$ approach to PMC one which has been done in \cite{mac4}. Some
other works which are related to PMC approach can be found in \cite{other1,other2}.

Converting the CORGI approach to the PMC one, is the other  research task which is not easy to do it and we hope to
be
able to report about it in future. The McKeon   ${\it et\; al.}$ approach can be considered at infrared fixed point where QCD $\beta$-function is
considered to be
zero \cite{fix33,fix40,fix42}. This subject is also interesting to be investigated from the other literature points of view like the infrared safe mode
of QCD  \cite{mine-preparation}  or the Ads/CFT   \cite{Brodsky:2010ur}  and would be valuable to follow it as our
further research activity.
\section*{Acknowledgment}
{{}Authors acknowledge Yazd university for  provided  facilities to do this project. Authors are indebted M.~R.~Khellat for his reading of the manuscript and constructive comments.}

\end{document}